\documentstyle[epsfig,aps]{revtex}
\begin{document}

\title{ 
{\hfill {\normalsize Preprint ITP-02-15E} }
\newline
\newline
Thermal Dilepton Radiation from Finite Fireball }

\author{D.~Anchishkin$^{a,}$\cite{email1}, V.~Khryapa$^{a,}$\cite{email2},
and V.~Ruuskanen$^{b,}$\cite{email3}}

\address{$^a$Bogolyubov Institute for Theoretical Physics,
          03143 Kiev, Ukraine}
\address{$^b$University of Jyv\"askyl\"a
P.O.Box 35, FIN-40351 Jyv\"askyl\"a, Finland   }

\date{\today}

\maketitle

\begin{abstract}

We analyze the dilepton emission rates from the hot pion gas confined to
a finite space-time volume.
Two models of $\pi^+\, \pi^-$ annihilation to dileptons are under
consideration.
The first model mimics an extreme influence of the dense hadron environment
on the $\rho$-meson mean free path and mean life time.
The second model deals with the standard concept of vector meson
dominance when the hadron medium is not accounted.
Our results indicate that the dilepton rates are finite in the low 
invariant mass region $M\le 2m_\pi$.
It is found that the rates experience an enhancement in the $e^+\, e^-$ 
production when invariant mass tends to the "real" threshold $M=2m_e$.
The breaking of the detailled energy-momentum conservation in favour of the 
integral one due to broken translation invariance is discussed.

\end{abstract}

\pacs{PACS numbers: 25.75.-q, 25.75.Gz, 03.75.-b}


\section{ Introduction}


Dilepton production in high energy nucleus-nucleus collisions provides an
information on the early high temperature and density stage where QGP
formation is expected.
Registration of dileptons gives important observables which probe the pion
dynamics in the dense nuclear matter that exists in the early stage of the
collision. 
After the creation, $e^+\, e^-$ and $\mu^+\, \mu^-$ pairs do not practically 
interact with the surrounding hadron matter and the analysis of the
final dilepton spectra experimentally obtained provides an excellent
possibility to investigate hadron fireball in its initial state.
The enhancement in the production of dileptons with an invariant mass
of 200-800 MeV observed by the CERES collaboration
\cite{agakichev95,agakichev98} has received considerable attention
recently and has been studied in the framework of various theoretical models.
It was found  that the most prominent channel of dilepton production
accounting a large part of the observed enhancement is pion annihilation
\cite{koch96,haglin}. 
The sensitivity of the dilepton spectra to a large variety of initial 
conditions for the hadron fireball was investigated.
Namely, the enhancement of dilepton production was explained  by 
modifications of pions in the medium and the pion-nuclear form-factors 
in the pion gas \cite{koch96}, or by the pion dispersion relation modified 
in the medium \cite{rapp96,song}.
The inclusion of the effects of baryon resonances \cite{rapp97,rapp99},
which couple directly to $\rho$-mesons, appear to be able to substantially 
increase the yield, although the calculation in \cite{steele} found a much
smaller effect due to baryons.
In Ref.~\cite{kluger}, the dilepton production was calculated as a result of
the interaction between long-wavelength pion oscillations or disoriental
chiral condensates and the thermal environment within linear sigma model.
It was shown that the dilepton yield with the invariant mass near and below
$2m _\pi$ can be up to two orders of magnitude
larger than the corresponding equilibrium yield due to soft pion modes.
Finite pion width effects and their influence on $\rho$-meson properties
and, subconsequently, on dilepton spectra were investigated in
\cite{van-hees2000}.
Whereas, the influence of non-equilibrium processes and finite times
were investigated in \cite{cooper}.
The conclusion of many works is that in order to better fit the eperimental 
data, an additional modification of the hadron phase in the medium and 
nonequilibrium processes (or effects) need to be considered.




The purpose of the present paper is to revise (refine) the
$\pi^+\, \pi^-$ annihilation mechanism of dilepton production from the
pion gas when this gas is confined to a finite space-time region.
The standard consideration of the reaction 
$ \pi^+\, \pi^- \to  \rho \to \gamma^* \to l\, \overline{l} $
\cite{gale87,mclerran85,domokos,ruuskanen,kapusta89} (see Appendix~A)
assumes an infinite space-time volume of the multipion system.
This immediately results in a "sharp" energy-momentum conservation
and presence of the $\delta$-function $\delta^4(K - P)$ as a factor of 
the S-matrix element (here $K=k_1+k_2$ is the total momentum of the 
pion pair and $P=p_+ +p_-$ is the total momentum of the lepton pair).
As a consequence of the equality $K=P$
the invariant mass of the two-pion system is specified as $K^2=M^2$,
where $M$ is the invariant mass of dilepton system
(i.e. $P^2=M^2$), which is a measurable quantity.
However, if the pion gas does not spread in all the space, but it occupies 
a finite space-time volume, and its life time is not long enough,
the $\delta$-function should be smeared to another distribution,
e.g. $\rho(K - P)$, which represents the distribution of the two-pion total
energy $K^0$ and total momentum ${\bf K}$ around the measurable quantities
$P^0$ and ${\bf P}$, respectively.
Effectively, this means the smearing of the pion pair invariant 
mass $M'$ around the mass $M$ of the lepton pair.
For instance, in case of a finite time interval,
one can use the Breit-Wigner energy distribution in place of
$\delta(K^0 - P^0)$.
On the other hand, in case of a
finite space volume, one can put a relevant form-factor
$\rho( {\bf K}- {\bf P} )$ instead of
$ \delta^3( {\bf K}- {\bf P} ) $
(we note, that these examples are appropriate for the non-relativistic 
sector).
The consequence of smearing the distribution in 4-momentum space is 
strong enough.
The rate becomes measurable under the standard threshold which 
is twice as the pion mass, i.e. for the values $M < 2m_\pi$, and down 
to the value of two lepton masses.
For instance, for the electron-positron pair production, the rate becomes 
measurable even in the vicinity of $2m_e\approx 1$~MeV, where $m_e$ is 
the electron mass.
Obviously, this phenomenon is due to the uncertainty principle.
In fact, if the life-time of the pion system is restricted, for instance by
$\tau =2\, $fm/c, then the energy (invariant mass) uncertainty is in the
range up to $\Delta E=100\, $MeV.
Actually, this means that lepton pairs with invariant masses down to
$M \propto 2m_\pi-100$~MeV can be registered.
Indeed, in accordance with quantum mechanics, we can speak about this matter
because the experimental data are taken from every particular event of
the collision of two relativistic ions.
As known, the strong interaction reactions (quark-antiquark annihilation,
pion-pion annihilation, etc) which are attributed to this event
take place during a finite time ($\tau=4--10~$fm/c) and in a restricted 
volume ($R=4--10~$fm).
The products of these reactions (secondary photons, lepton pairs, etc.)
are registered and measurements can be distinguished from event to event.
That is why, it is reasonable to consider a finite life time and a finite
spatial volume occupied by the system of particles created
in relativistic heavy ion collisions when every event  measurements
take place.

One another important consequence of a finiteness of reaction space-time
volume 
is a cutting of "exact connection" between the lepton-lepton c.m.s. 
and pion-pion c.m.s.
Indeed, the relation between the total momentum $P$ of an outcoming lepton
pair and the total momentum $K$ of an incoming pion pair (see Fig.~1) is
weighted now by the distribution function $\rho(K - P)$, which can be regarded
as a smeared pattern of $\delta(K - P)$.
That is why, any quantity which is determined (evaluated) in the lepton-lepton
c.m.s., which moves with velocity
$ {\bf v}_P = {\bf P}/P_0 $ in the lab system, should be Lorentz transformed
to  the pion-pion c.m.s., which in turn moves with velocity
${\bf v}_K= {\bf K}/K_0$ in the lab system.
As we shall see, such a transformation increases the electron-positron 
emission rate for small invariant masses $2m_e < M < 2m_\pi$.

Actually, a restriction of the space-time volume in the process
$a\, \bar{a} \to b\, \bar{b}$ induces several effects which should be
taken into account.
Basically, the nonequilibrium behavior results in a correction to the
standard spectrum obtained in the infinite space-time.
On the other hand, this correction can be
conventionally separated into three different contributions:
a)~First, it is a nonstationary behavior of the many-particle system.
If, for instance, the multipion system which provides pions for
the process
$\pi^+\, \pi^- \to l\, \bar{l}$ has finite life time, then individual
pion states are also nonstationary and a decay of the states should be taken
into account.
b) The second contribution comes from the direct presence of the form-factor
of the multipion system which practically determines a finiteness of
space-time volume.
c)
And at least, as we just discussed above, the third contribution comes from
smearing 
the $\delta$-function connection between the total 4-momentum of incoming
particles (pions) and the total 4-momentum of outcoming particles (leptons).

The goal of the present paper is to estimate the influence of the finite
space-time volume of the hadron reaction zone on the dilepton emission rate
taking into account the listed above  second and third contributions.
We do not consider a nonstationarity of annihilating individual pion states
and reserve investigation of this problem to the next paper.


\section{Two-pion annihilation in finite space-time volume}


\medskip

We start with a simple model:
a) A thermalized hadron matter which is characterized by the temperature $T$,
mean life time $\tau$, and mean finite volume $V$ is created in a central
heavy ion collision;
b) The pion gas created in the collision can be regarded in the first
approximation
as an ideal one with single-particle distribution function $f_{\rm th}(E)$.

The amplitude of the pion-pion annihilation reaction
in the first nonvanishing approximation (see Fig.~1a, 2b) reads as:
pions annihilate through the $\rho $-meson form-factor (vertex), then
in accordance with vector dominance through a virtual photon which creates
a lepton-antilepton pair, i.e.
$ \pi^{+}\, \pi^{-} \to \rho \to \gamma^{*} \to l\, \overline{l} $.
For the sake of transparency of the effects under investigation, we write
the amplitude in semi-naive way ({\it a la} Bjorken, Drell, V.I).
To reflect the finiteness of the space-time volume of the pion gas, we 
introduce the pion 4-density $\rho (x)$, so that the vertex $x_1$ in 
Fig.~1a, 2b is weighted by this density.
Then, the amplitude of the process reads
\begin{figure}[t]
\begin{center}
\epsfig{file=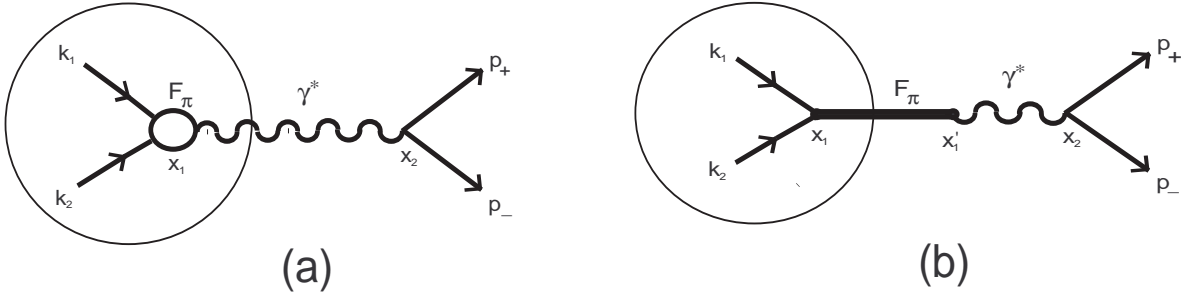,height=4.5cm,width=18cm,angle=0}
\caption{ The first order nonvanishing Feynman graph
of the lepton pair creation  in the process
$ \pi^{+}\, \pi^{-} \to \rho \to \gamma^{*} \to  l\, \overline{l} $.
The circle which surrounds the vertex $x_1$ sketches the finite
space-time region of the pion-pion interactions.
  }
\vspace{-0.2cm}
\end{center}
\end{figure}
\noindent
\begin{eqnarray}
A_{fi}\left( k_1,k_2;p_+,p_- \right)
=&&
4\pi \alpha \int \rho(x_1) d^4 x_1 \, \int d^4 x_2 \,
\phi ^{*}_{k_1}(x_1) \phi ^{*}_{k_2}(x_1)\,
F_{\pi }\left( (k_1+k_2)^2 \right) \,
D_{F}(x_1-x_2)
\nonumber \\
&& \! \times \, \phi _{p_+}(x_2)\, \phi _{p_-}(x_2) \,
m_{fi}\left( k_1,k_2;p_{+},p_{-} \right)
\  ,
\label{2-1}
\end{eqnarray}
where $\alpha =e^2/ 4\pi$,
$\phi_k (x)=[(2\pi )^32E_k]^{-1/2}\exp{(-ik\cdot x)} $
are the initial pion states with $k=k_1$, $k=k_2$ and the final lepton states
with $k=p_{+}$, $k=p_{-}$ (lepton spinor part is hidden in the matrix
element $m_{fi}$).
The on-shell energy of the incoming and outgoing particles are defined as:
$k^0_i=E_i=\sqrt{m_\pi^2+k_i^2}$ with $i=1,2$ for pions and
$p^0_{\pm}=E_{\pm}=\sqrt{m_e^2+p_{\pm}^2}$ for leptons, respectively.
The quantity $F_\pi(K)$ is the $\rho $-meson form factor (\ref{3});
$D_F(x_1-x_2)$ is the photon propagator, and $m_{fi}$ is the matrix element
\begin{eqnarray}
| m_{fi}|^2 &=&
\sum_{spins} t_\nu^+ t_\mu (k_1-k_2)^\nu (k_1-k_2)^\mu
\nonumber \\
&=&  8(k_1-k_2)\cdot p_{+}\
(k_1-k_2)\cdot p_- - 4(k_1-k_2)^2(m_e^2+p_+\cdot p_-)
\ ,
\label{2-2}
\end{eqnarray}
where $t_\nu = \bar{u}(p_-)\gamma_\nu u(p_+)$ with $u(p)$ as electron spinor.
In the momentum space representation, amplitude (\ref{2-1}) looks like
\begin{equation}
A_{\rm fi}\left( k_1,k_2;p_+,p_- \right)
=
\frac{\displaystyle \alpha F_{\pi }\left( (k_1+k_2)^2 \right)}
              {\displaystyle 2\pi^2 \sqrt{2E_{1}2E_{2} }}
         \rho (k_1+k_2-p_+ - p_- ) D_F(p_{+} + p_{-})
         \frac{m_{fi}(k_1,k_2;p_{+},p_{-})}
              {\displaystyle (2\pi )^{3}\sqrt{2E_{+}2E_{-} }} \ ,
\label{2-3}
\end{equation}
where
$\rho (k)=\int d^{4}x~\rho (x)\exp{(ik\cdot x)} $
is the Fourier transform of the pion density (pion source form-factor) 
and it obviously stands in place of the delta function
$(2\pi)^4 \delta^4(k_1+k_2-p_+ - p_- )$
due to the finite space-time volume occupied by the pion system which 
was created during a nucleus-nucleus collision.

It should be pointed out that we consider two models (a) and (b) of
the pion-pion annihilation.
Each of the models can be putted in correspondence to Fig.~1a or to Fig.~1b.
Model (b) (Fig.~1b) reflects a standard representation of the concept of
vector meson dominance where, for the description of $\rho$-meson 
propagation, we do not take into account the hadron environment.
Whereas, model (a) depicted in Fig.~1a accumulates, in some sense, 
an extreme distortion of the $\rho$-meson behavior due to the dense hadron
environment.
Indeed, due to the environment, $\rho$-mesons experience a suppression
of the mean free path and mean life time.
If one assumes an increase of the $\rho$-meson width
$\Gamma_\rho$ due to the hadron medium, then the $\rho$-meson life time
is comparable to the life time of the pion subsystem and, hence, the mean 
free path is confined to the fireball volume.
Actually, as the limit of the decreasing $\rho$-meson
mean free path, one can take a zero path.
Following this idea, model (a) implies a point-like decay of a
$\rho$-meson to a photon $\gamma^*$.
Meanwhile, to preserve a $\rho$-meson nature of the pion-pion annihilation,
we use the $\rho$-meson form-factor $F_{\pi }(K)$  (\ref{3}) with the vacuum
values of the parameters.

Amplitude (\ref{2-1}) was written in accordance with model (a).
On the other hand, in the standard vector-meson dominance treatment, which is
reflected by Fig~2b, the explicit propagation of a $\rho$-meson is taking
into account.
The formal difference of these two approaches comes out just as the
difference of the arguments of the function
$F_\pi(k^2)$ ($\rho$-meson propagator), which for model (a) is
$F_\pi\left( (k_1+k_2)^2 \right)$ (Fig~2a), 
whereas it is $F_{\pi }\left( (p_++p_-)^2 \right)$ for model (b) (Fig~2b).
In the momentum space representation, the amplitude which corresponds 
to the diagram depicted in Fig.~1b reads
\begin{equation}
A_{\rm fi}\left( k_1,k_2;p_+,p_- \right)
=
\frac{\displaystyle \alpha F_{\pi }\left( (p_{+} + p_{-})^2 \right)}
              {\displaystyle 2\pi^2 \sqrt{2E_{1}2E_{2} }}
         \rho (k_1+k_2-p_+ - p_- ) D_F(p_{+} + p_{-})
         \frac{m_{fi}(k_1,k_2;p_{+},p_{-})}
              {\displaystyle (2\pi )^{3}\sqrt{2E_{+}2E_{-} }} 
\ .
\label{2-3a}
\end{equation}
It is interesting to note that these two models, (a) and (b), coincide with 
each other when there is no restriction on the 4-volume of the reaction zone.
But it is not the case when we deal with the finite space-time volume of the
reaction.
We will analyze the consequences of the difference of these two
approaches on the rate of dilepton emission.

The number of dileptons produced per event in the element of the 4-momentum
space $d^4P$ reads
\begin{equation}
\frac{\displaystyle dN^{(\rho)} }
{\displaystyle d^3k_1\, d^3k_2\, d^4P} =
\int d^3p_{+}\, d^3p_{-}~\left| A_{\rm fi} \left( k_1,k_2;p_+,p_- \right)
\right|^2 \,
\delta^4(p_{+}+p_{-}-P)
\ ,
\label{2-4}
\end{equation}
where the super index $(\rho)$ denotes the presence of the form-factor 
$\rho(K-P)$ or finite pion 4-density $\rho(x)$.
Taking an explicit form of amplitude (\ref{2-3}), the expression on the
r.h.s. of (\ref{2-4}) can be represented in the following form:
\begin{equation}
\frac{ \displaystyle dN^{(\rho)} }
{ \displaystyle d^3 k_1 d^3 k_2 d^4P }
=
\left| \rho (k_1+k_2-P) \right| ^2 \,
\frac{\displaystyle dN}
{\displaystyle d^3 k_1 d^3 k_2 d^4P} \ ,
\label{2-5}
\end{equation}
where $dN/d^3k_1d^3k_2d^4P$ is the "standard" rate of particle pair
production with the total momentum $P$,
\begin{equation}
\frac{\displaystyle dN}
{\displaystyle d^3k_1\, d^3k_2\, d^4P} =
\frac{\displaystyle \alpha^2\, |F_\pi\left( (k_1+k_2)^2 \right)|^2}
              {\displaystyle 4\pi^4\, 2E_1\, 2E_2\, P^4 }
\int \frac{d^3p_{+}}{(2\pi )^3\, 2E_{+} }\,
     \frac{d^3p_{-}}{(2\pi )^3\, 2E_{-} }\,
     ~\left| m_{fi}\right| ^{2} \delta^4(p_{+}+p_{-}-P)
\ .
\label{2-6}
\end{equation}
Meanwhile, there are new important features relevant to (\ref{2-6})
which are the consequence of the space-time restrictions on the
vertex $x_1$ in the diagrams depicted in Fig.~1.
Indeed, when we evaluate the expression on the r.h.s. of (\ref{2-6}),
we should take into account that the pion-pion c.m.s. moves in the lab
system with velocity
${\bf v}_K= {\bf K}/K_0$, where $K=k_1+k_2$,
whereas the dilepton c.m.s. moves with velocity
${\bf v}_P= {\bf P}/P_0$ in the lab system.
These two center-off-mass systems are "disconnected" now and the relation
between the total momenta $K$ and $P$ is weighted by the form-factor
$\left| \rho (K-P) \right| ^2$
which stands as a prefactor on the r.h.s. of (\ref{2-5}).
That is why, the velocities ${\bf v}_K$ and ${\bf v}_P$ do not equal each
other as it obviously was in the standard approach which deals with
the infinite space-time.

Actually, we meet here the integral energy-momentum conservation which is
guaranteed by relation
$\int \rho (K-P) d^4 K/(2\pi^4) =1$ instead of the detailled conservation,
which was represented by $\delta^4(K-P)$  when the
space-time interval is infinite.
This phenomenon is typical in quantum physics when we deal with finite
time intervals (energy uncertainty) or finite coordinate space intervals
(momentum uncertainty).
In essence, any system, for instance with finite life-time, possesses
a resonance-like
behavior and, hence, the energy-smearing function $\rho (K^0-P^0)$ can 
look like the
Breit-Wigner function (Lorentz shape) or a similar one, for example
$\rho(K^0)=\frac{1}{\pi}\, \frac{\Gamma/2}{(K^0-P^0)^2-\Gamma^2/4}$
with $\Gamma=1/\tau$ as the width of the energy probability distribution
($\tau$ is the life time).
In this probability distribution, $P^0$ represents a mean value because it
is the measured (external) quantity and $K^0$ is a current value of energy
which distributed around $P^0$.
In our case, the total energy of the pion pair $K^0$ is distributed around
a fixed value $P^0$ (dilepton total energy) and this is a consequence of
the finiteness of the pion gas life time.

Putting the result of evaluation of the integral on the r.h.s. of (\ref{2-6}) 
into (\ref{2-5}), we obtain (we skip over details of this evaluation)
\begin{equation}
\frac{ \displaystyle dN^{(\rho)} }{ \displaystyle d^3k_1\, d^3k_2\, d^4P }
=
\left| \rho (k_1+k_2-P) \right| ^2 \,
\frac{\displaystyle \alpha^2\,
\left| F_\pi \left( (k_1+k_2)^2 \right) \right|^2 }
              {\displaystyle 4\pi^4\, 2E_1\, 2E_2\, P^2 }
\frac{ \left( {\bf k}_1-{\bf k}_2 \right)_P^2 }{ 3(2\pi)^5 } \,
\left( 1+\frac{2m_e^2}{P^2} \right)
\left( 1-\frac{4m_e^2}{P^2} \right)^{1/2}
\ ,
\label{2-7}
\end{equation}
where integration was done in the lepton pair c.m.s. 
Subindex $P$ denotes that the quantity
$ ({\bf k}_1-{\bf k}_2)^2 $ is taken in the P-system, by definition it is
lepton pair c.m.s.
Whereas, subindex $K$ denotes that the quantity is taken in the K-system, 
which is the pion pair c.m.s.
As we discussed above, such a separation is due to the fact that velocity
${\bf v}_K$ of the K-system and velocity ${\bf v}_P$ of the P-system do 
not coincide with each other.
Further, the difference $ ({\bf k}_1-{\bf k}_2)_P^2 $ will be 
Lorentz-transformed to $ ({\bf k}_1-{\bf k}_2)_K^2 $ (see Appendix~B).


The next step of evaluations needs the averaging of the quantity
$ dN^{(\rho)}/d^3k_1\, d^3k_2\, d^4P $ (\ref{2-7}) over the pion
momentum space with the distribution function which reflects the model 
of a pion system created in the relativistic heavy ion collision.
For our estimation, we take the widely used model where the source is
nothing more as a pion gas which is in the thermal equilibrium
characterized by the temperature $T$.
Then, one has to weight the incoming pion momenta by a thermal
distribution function $f_{\rm th}(E)$ and make integration with respect 
to these momenta
\begin{equation}
\left\langle
\frac{ \displaystyle dN^{(\rho)} }{ \displaystyle d^4 P }
\right\rangle
=
\int d^3k_1~f_{\rm th}(E_1)\int d^3k_2~f_{\rm th}(E_2)
\frac{ \displaystyle dN^{(\rho)} }
     { \displaystyle d^3k_1d^3k_2d^4P }
\ .
\label{2-8}
\end{equation}
It is pedagogical to represent this equation in the form where explicit
factorization of the pion source form-factor $\rho(K-P)$ is made:
\begin{eqnarray}
\left \langle
\frac{ \displaystyle dN^{(\rho)} }{ \displaystyle d^4P }
\right \rangle
=
\int d^4 K~\left| \rho (K-P)\right|^2
\left \langle
\frac{\displaystyle dN}{\displaystyle d^4 K~d^4 P}
\right \rangle
\ .
\label{2-9}
\end{eqnarray}
Here, we introduce the auxilliary quantity
\begin{equation}
\left \langle
\frac{\displaystyle dN}{\displaystyle d^4K~d^4P}
\right \rangle
\equiv
\int d^3 k_1~f_{\rm th}(E_1)\int d^3 k_2~f_{\rm th}(E_2)
\frac{\displaystyle dN}{\displaystyle d^3 k_1 d^3 k_2 d^4P}
\delta^4 (k_1+k_2-K)
\, ,
\label{2-10}
\end{equation}
which is the averaged dilepton number density with respect to the total
4-momenta $K$ and $P$.
In this formula the spectrum $dN/d^3 k_1 d^3 k_2 d^4P$ equals to expression 
on the r.h.s. of (\ref{2-7}) without the first factor 
$\left| \rho (k_1+k_2-P) \right|^2$.
The meaning of the averaging on the r.h.s. of (\ref{2-10}) is quite
transparent: we integrate over pion momenta with weights which are the
distribution functions of pions in the momentum space, keeping meanwhile
the total energy-momentum of the pion pair as a constant equal to $K$.
The resulting quantity on the l.h.s. of (\ref{2-10}) looks like we cut
the s-channel of the reaction $\pi^+\pi^- \to l\bar{l}$ into two independent
blocks (vertices), see diagrams depicted in Fig.~1, and calculate
the distribution with respect
to the incoming total momentum $K$ and outgoing total momentum $P$ independently.
Then, equation (\ref{2-9}) is the connection of these vertices with a
weight function, the form-factor $ \left| \rho (K-P)\right|^2 $, reflecting
that the pion gas lives in a finite space-time volume, which results in 
the resonance-like behavior of the pion pair.
On the other hand, the form-factor $ \left| \rho (K-P)\right|^2 $ on the r.h.s. of
(\ref{2-9}) can be regarded as a distribution function which is centered
around a mean value $P$ (because it is measured or externally fixed quantity) and
the averaging integration is carried on over the fluctuating quantity $K$.

Taking into account a result of integration of the expression on
the r.h.s. of (\ref{2-10}) (we skip over details of this evaluation),
we can rewrite (\ref{2-9}) in the following form:
\begin{eqnarray}
\left\langle
\frac{ \displaystyle dN^{(\rho)} }{ \displaystyle d^4P }
\right\rangle
=&&
\frac{ \displaystyle \alpha^2 C^2 }{ \displaystyle 3(2\pi )^8 }
\left( 1- \frac{4m_e^2}{P^2} \right)^{1/2}
\left( 1+ \frac{2m_e^2}{P^2} \right)\,
\int d^4 K \left| \rho (K-P) \right|^2
e^{-\beta u\cdot K} \frac{K^2}{P^2}\,
\big| F_\pi(K^2) \big|^2
\left( 1- \frac{4m_\pi^2}{K^2} \right)^{3/2}
\nonumber \\
&& \! \times \
\left[
1 + \frac{1}{3} \left( \frac{ \left(P\cdot K\right)^2 }{P^2\, K^2} -1 \right)
\right]
\ ,
\label{2-12}
\end{eqnarray}
where $u$ is the 4-velocity of the element of the fireball which is in
quasiequilibrium state.
We take the Boltzmann distribution
$f_{\rm th}(E)=C\exp (-\beta E)$ as a pion thermal distribution function 
with inverse temperature $\beta =1/T$ and $C$ to be a normalization constant.

The last factor in square brackets on the r.h.s. of (\ref{2-12}) is a
correction which is due to the Lorentz transformation of the quantity
$\left( {\bf k}_1-{\bf k}_2 \right)^2$ from the P-system to the K-system
(see Appendix~B).
We need to do the Lorentz transformation because the integration (thermal 
averaging (\ref{2-10})) at this time was done in the pion pair c.m.s.
The factor $1/3$ in the brackets occurs after the integration over angles.

Expression (\ref{2-12}) is a final result for the distribution
of the number of particles as a function of dilepton 4-momentum $P$.
Further evaluation can be made for a particular form
of the pion gas form-factor $\rho (K-P)$.

Let us add for completeness, that for the process which corresponds to
the diagram depicted in Fig.~1b (model (b)), one should correct the
last expression (\ref{2-12}) by shifting the pion form-factor, which is
now taken in the form
$\big| F_\pi(P^2) \big|^2$,
in front of the integral.

The convergence of (\ref{2-12}) to the standard result can be recognized when
one changes the pion gas form-factor to the $\delta$-function, namely
$\rho (K-P) \to (2\pi)^4 \delta(K-P)$.
It is easily seen that the expression in square brackets immediately 
transforms to unity when $K=P$ because there is no more disconnection 
between the P- and the K-systems. 
They both move now with velocity
${\bf v}_K={\bf v}_P={\bf P}/P_0$ in the lab system.
We interpret, as usual, one power of the $\delta$-function as 4-volume, i.e.
$ (2\pi)^4 \delta^4(K-P) = \int d^4x \exp{ [-i(K-P)\cdot x] }|_{K=P}=\tilde TV $,
where $\tilde T$ is the time interval and $V$ is the volume.
Then, dividing by the 4-volume $\tilde TV$, we obtain the rate 
($R\equiv N/\tilde TV$)
\begin{eqnarray}
\frac{1}{T\, V} \,
\left\langle
\frac{ \displaystyle dN^{(\rho)} }{ \displaystyle d^4P }
\right\rangle \
\to \
\left\langle
\frac{ \displaystyle dR }{ \displaystyle d^4P }
\right\rangle
=
\frac{ \displaystyle \alpha^2 C^2 }{ \displaystyle 3(2\pi )^8 }\,
e^{-\beta P_0}
\big| F_\pi(P^2) \big|^2 \,
\left( 1- \frac{4m_\pi^2}{P^2} \right)^{3/2}
\left( 1- \frac{4m_e^2}{P^2} \right)^{1/2}
\left( 1+ \frac{2m_e^2}{P^2} \right)
\ .
\label{2-12a}
\end{eqnarray}
To obtain the distribution of the number of lepton pairs with respect
to the invariant mass $M$ of the pair it is necessary to integrate distribution
(\ref{2-12}),
$ \left\langle dN^{(\rho)}/d^4P \right\rangle $,
over the 4-momentum space $P$ setting the value $P^2$ to $M^2$, namely
\begin{eqnarray}
\left \langle
\frac{ \displaystyle dN^{(\rho)} }{ \displaystyle dM^2 }
\right \rangle
&=&
\int d^4 P \,
\left \langle
\frac{ \displaystyle dN^{(\rho)} }{ \displaystyle d^4P }
\right \rangle \,
\delta ( P^2 - M^2 )\, \theta (P_0)
\nonumber \\
&=&
\int \frac{d^3 P}{ 2\sqrt{M^2+{\bf P}^2} } \,
\left \langle
\frac{ \displaystyle dN^{(\rho)} }{ \displaystyle d^4P }
\right \rangle \bigg|_{P_0=\sqrt{M^2+{\bf P}^2}} \,
\ .
\label{2-14}
\end{eqnarray}
Hence, taking our result (\ref{2-12}), we can write the distribution of
the number of created lepton pairs $N^{(\rho)}$ with respect to the invariant 
mass of a lepton pair $M$
\begin{eqnarray}
\left\langle
\frac{ \displaystyle dN^{(\rho)} }{ \displaystyle dM^2 }
\right\rangle
=&&
\frac{ \displaystyle \alpha^2 C^2 }{ \displaystyle 3(2\pi )^8 }
\left( 1- \frac{4m_e^2}{M^2} \right)^{1/2} \!
\left( 1+ \frac{2m_e^2}{M^2} \right)
\int \! \frac{d^3 P}{ 2\sqrt{M^2+{\bf P}^2} } \, d^4 K
\left| \rho (K-P) \right|^2
e^{-\beta K_0} \frac{K^2}{M^2}\,
\big| F_\pi(K^2) \big|^2
\left( 1- \frac{4m_\pi^2}{K^2} \right)^{3/2}
\nonumber \\
&& \! \times \
\left[
1 + \frac{1}{3} \left( \frac{ \left(P\cdot K\right)^2 }{M^2\, K^2} -1 \right)
\right] \,
\theta\left(K_0\right) \, \theta\left(K^2-4m_\pi^2\right)
\bigg|_{P_0=\sqrt{M^2+{\bf P}^2}}
\ ,
\label{2-15a}
\end{eqnarray}
where, by the presence of the $\theta$-functions (last two factors on the
r.h.s. of (\ref{2-15a})), we would like to stress that the invariant mass
of a pion pair $M_\pi^2 \equiv K^2$ is not less than two pion masses.
On the other hand, possible finite values of the distribution
$ \left\langle dN^{(\rho)}/dM^2 \right\rangle $ below the two pion mass
threshold can occur just due to the presence of the many-particle pion system
form-factor $\rho (K-P)$.

Note, if one follows model (b) which corresponds to the process
depicted in Fig.~1b, then the pion form-factor on the r.h.s. of (\ref{2-15a})
should be taken in the form
$ \big| F_\pi(M^2) \big|^2 $
and shifted to the front of the integral.

Equation (\ref{2-15a}) is the main theoretical result of the
present paper.
The next steps in the investigations of the subject can be made by taking
particular form-factor $\rho (K-P)$ of the many-particle (multipion)
system.


\subsection{ Gaussian pion gas form-factor }


To make an access to the effects under investigation transparent as much as
possible, we take, as a model of the pion source, the Gaussian distribution
in space and the Gaussian decay of the system of interacting pions
\cite{yano}
\begin{equation}
\rho (x) =
\exp{ \left( - \frac{t^2}{2\tau^2} - \frac{{\bf r}^2}{2R^2}  \right) }
\ ,
\label{2-14a}
\end{equation}
which transforms to $1$ for the large enough mean life time $\tau$ and mean
radius $R$ of the system, i.e. when $\tau \to \infty,\, R \to \infty$,
$\lim \rho (x) = 1 $ and amplitude (\ref{2-1}) comes to the standard form.
The Fourier transformation of
$ \rho (x) $ gives the form-factor of the multipion system as
\begin{equation}
\rho (Q)
=
 (2\pi)^2 \tau R^3 \exp{\left[-\frac{1}{2} \left(Q_0^2\tau^2
 +
 {\bf Q}^2R^2 \right) \right] }
\ ,
\label{2-15}
\end{equation}
We should also keep correspondence with the form-factor for the infinite
space-time volume, which is nothing more as a $\delta$-function times
$(2\pi)^4$.
But the correspondence should be installed for the second power of
form-factors
\begin{equation}
\lim _{\tau,\, R \to \infty }\rho^2 (Q) =T V~(2\pi )^4\delta^4 (Q)
\ ,
\label{2-16}
\end{equation}
where, on the r.h.s., we transformed one $\delta$-function to the 4-volume,
$\tilde T$ is a time interval and $V$ is a spatial volume.
On the other hand, for the l.h.s. of (\ref{2-16}), we have explicitly
\begin{equation}
\left( \pi^2 \tau R^3 \right) \,
(2\pi)^4
\lim_{\tau \to \infty }
\frac{\tau}{\pi^{1/2}} e^{-Q_0^2\tau^2}\
\lim_{R \to \infty }
\frac{R^3}{\pi^{3/2}} e^{-{\bf Q}^2R^2}
=
\left( \pi^2 \tau R^3 \right) \,
(2\pi)^4 \delta(Q_0)  \delta^3({\bf Q})
\ .
\label{2-16a}
\end{equation}
Hence, putting the r.h.s. of (\ref{2-16}) and the r.h.s. of (\ref{2-16a}) 
in correspondence, we can conclude that the quantity $\pi^2 \tau R^3 $ 
should represent the 4-volume $\tilde TV$ the for source parametrization
(\ref{2-14a}).

With the use of the above source distribution function, Eq.~(\ref{2-15a}) 
yields the spectrum with respect to lepton
pair invariant mass
\begin{eqnarray}
\frac{1}{\pi^2 \tau R^3} \,
\left\langle
\frac{ \displaystyle dN^{(\rho)} }{ \displaystyle dM^2 }
\right\rangle
=
\frac{ \displaystyle \alpha^2 C^2 }{ \displaystyle 3(2\pi )^8 }
\left( 1- \frac{4m_e^2}{M^2} \right)^{1/2} \!
\left( 1+ \frac{2m_e^2}{M^2} \right)
\int d^4 K \,
\theta\left(K_0\right) \, \theta\left(K^2-4m_\pi^2\right) \,
e^{-\beta K_0} \frac{K^2}{M^2}\,
\big| F_\pi(K^2) \big|^2 \,
\nonumber \\
 \times \,
\left( 1- \frac{4m_\pi^2}{K^2} \right)^{3/2}
\int \frac{d^3 P}{ 2\sqrt{M^2+{\bf P}^2} } \,
(2\pi)^4
\frac{\tau}{\pi^{1/2}} e^{-\left( K_0-P_0 \right)^2 \tau^2}\,
\frac{R^3}{\pi^{3/2}} e^{-\left( {\bf K}-{\bf P} \right)^2 R^2}
\left[
1 + \frac{1}{3} \left( \frac{ \left(P\cdot K\right)^2 }{M^2\, K^2} -1 \right)
\right]
\ ,
\label{2-17}
\end{eqnarray}
where
$P_0=\Omega({\bf P}) \equiv \sqrt{M^2+{\bf P}^2} $.
We write the source form-factor exponents together with relevant
coefficients in
the form which shows the evident transformation to the $\delta$-functions
$ \delta\left( K_0-P_0 \right) $
and
$ \delta^3\left( {\bf K}-{\bf P} \right) $
when $\tau \to \infty$ and $ R \to \infty $, respectively.

After the integration over angle variables 
in both integrals on the r.h.s. of Eq.~(\ref{2-17}), we obtain the rate
($R^{\rm (\rho)}\equiv N^{(\rho)}/(T\, V)$) of dilepton production
in the reaction
$\pi^+\, \pi^- \rightarrow l\, \overline{l}$
\begin{eqnarray}
\left \langle
\frac{ \displaystyle dR^{\rm (\rho)} }{ \displaystyle dM^2 }
        \right \rangle
=
\frac{\displaystyle \alpha ^{2}C^{2} }
              {\displaystyle 3(2\pi )^3 }
\left( 1- \frac{4m_e^2}{M^2} \right)^{1/2} \! \!
\left( 1+ \frac{2m_e^2}{M^2} \right)
\int _{-\infty }^{\infty } \frac{\bar{P} d\bar{P} }{P_0}
\int_0^{\infty } \bar{K} d\bar{K}
~\frac{R}{\pi ^{1/2}}
e^{-(\bar{K} -\bar{P} )^2 R^2 }
\int _{ \sqrt{4m_{\pi }^2+\bar{K}^2} }^{\infty } dK_0 \
e^{-\beta K_0}
\nonumber \\
\times \
\frac{\tau }{\pi^{1/2}}
e^{ -(K_0 - P_0)^2\tau^2 }
\frac{K^2}{M^2}\,
\big| F_\pi(K^2) \big|^2 \,
\left( 1- \frac{4m_\pi^2}{K^2} \right)^{3/2}\,
\left[
     1+\frac{1}{3M^2 K^2}
     \left(
     \left( P\cdot K \right)^2
     - M^2 K^2
     +
     \frac{ P\cdot K }{ R^2 }
     +
     \frac{1}{ 2 R^4 }
     \right)
\right]
\ ,
\label{2-18}
\end{eqnarray}
where we use the notations
\begin{equation}
\bar{K} \equiv |{\bf K}| \ ,\ \ \bar{P} \equiv |{\bf P}| \ , \ \
P_0 = \Omega( \bar{P} ) = \sqrt{M^2+\bar{P}^2 } \ , \ \
K^2 = K_0^2-\bar{K}^2 \ , \ \
P\cdot K = \Omega( \bar{P} ) K_0 - \bar{P} \bar{K}
\, .
\label{2-19}
\end{equation}
Let us recall that the expression in square brackets on the r.h.s. of 
(\ref{2-18}) is
a relativistic correction which comes out when we transform vector
${\bf q}={\bf k}_1-{\bf k}_2$ from the lepton pair c.m.s. (P-system) to 
the pion pair c.m.s. (K-system).
This correction is due to the difference of velocities
${\bf v}_K={\bf K}/K_0$ and ${\bf v}_P={\bf P}/P_0$ of the K-system and 
the P-system in the lab system, respectively.
And in turn, the difference of velocities is caused by the pion
energy-momentum fluctuations which are due to the volume and life time
finiteness of the system.
For instance, the influence of the finite spatial size of the system on 
the relativistic correction discussed is represented now by two terms
$\propto R^{-2}$ and $\propto R^{-4}$ which explicitly appeared in
square brackets on the r.h.s. of (\ref{2-18}) and naturally reflect the
fact that for a smaller pion system, the relativistic correction is larger.
On the other hand, the appearance of these terms can be regarded as the
expansion over the inverse spatial size (scale) of the system.

Note again, that, for model (b) corresponding to the process
depicted in Fig.~1b, just one correction should be done in expressions
(\ref{2-17}) and (\ref{2-18}).
Formally, it is necessary to change the pion form-factor argument from
$K^2$ to $M^2$ and then it can be written in front of the integral
in the form
$\big| F_\pi(M^2) \big|^2$.


\subsection{Passage to the limits $R\to \infty$ and $\tau \to \infty$ }


First we consider the limit $R \rightarrow \infty $, what means that the
multipion system occupies the infinite volume in the coordinate space but lives
for a finite life-time in this state.
Using the limit relation
$ \lim_{R\to \infty} R \pi ^{-1/2}
\exp{[-(\bar{K}-\bar{P})^2 R^2]} = \delta(\bar{K}-\bar{P}) $,
one can perform integration in Eq.~(\ref{2-18}) with the use of the
$\delta$-function and obtain
\begin{eqnarray}
\left \langle
\frac{ \displaystyle dR^{\rm (\rho)} }{ \displaystyle dM^2 }
        \right \rangle_{R\to \infty}
=
\frac{\displaystyle \alpha ^{2}C^{2} }
              {\displaystyle 3(2\pi )^3 }
\left( 1- \frac{4m_e^2}{M^2} \right)^{1/2} \! \!
\left( 1+ \frac{2m_e^2}{M^2} \right)
\int _0^{\infty } \frac{\bar{K}^2 d\bar{K} }{\Omega(\bar{K})}
\int _{\sqrt{4m_{\pi }^2+\bar{K}^2} }^{\infty } dK_0 \
\frac{\tau }{\pi^{1/2}}
e^{ -[K_0 - \Omega(\bar{K})]^2\tau^2 }
e^{-\beta K_0}
\nonumber \\
\times \
\frac{K^2}{M^2}\,
\big| F_\pi(K^2) \big|^2 \,
\left( 1- \frac{4m_\pi^2}{K^2} \right)^{3/2}\,
\left[
     1+\frac{\bar{K}^2 \left( K_0 - \Omega(\bar{K}) \right)^2}{3M^2 K^2}
\right]
\ ,
\label{2-20}
\end{eqnarray}
where
$\Omega( \bar{K} ) = \sqrt{M^2+\bar{K}^2 } \, , \ \ K^2 = K_0^2-\bar{K}^2 $.
It is reasonable to note that, due to a finite life time $\tau$ of the
multipion system and, thus, a finite life time of annihilating pion states,
the off-shell behavior of the two-pion system is seen pretty well.
Indeed, this is reflected by the presence of the distribution
$e^{ -[K_0 - \Omega(\bar{K})]^2\tau^2 }$ of the pion pair total energy
$K_0$ around the nominal quantity
$\Omega(\bar{K})=\sqrt{ M^2+{\bf K}^2 }$ in the integrand on the r.h.s.
of (\ref{2-20}).
The same discrepancy of $K_0$ and $\Omega(\bar{K})$ determines the
relativistic correction (reflection of the difference in velocities of
the K-system and the P-system) which results in the presence of square brackets
on the r.h.s. of (\ref{2-20}).

Let us consider now another limit - we take
the infinite life-time of the source ($\tau \rightarrow \infty $)
but a finite volume of the reaction region.
This leads to the equality of energies or the equality of the zero components 
of the total momenta
$K_0$ and $P_0=\Omega(\bar{P})$ of the pion pair and lepton pair, respectively.
Indeed, taking the limit $\tau\to \infty$ on the r.h.s. of Eq.~(\ref{2-18}),
one gets
$\lim_{\tau\to \infty } \tau \pi^{-1/2}
\exp{ \left[-\left(K_0-\Omega(\bar{P}) \right)^2\tau^2\right] } =
\delta \left(K_0-\Omega(\bar{P}) \right)$.
It should be pointed that the result of integration over $K_0$ on the r.h.s.
of (\ref{2-18}) with the help of this $\delta$-function is not zero when
the following inequality takes place
\begin{eqnarray}
\sqrt{4m_{\pi }^2+\bar{K}^2} \le \Omega\left( \bar{P} \right)
\ .
\label{2-23}
\end{eqnarray}
If the measured dilepton invariant mass $M$ is larger than $2m_\pi$,
inequality (\ref{2-23}) results in a restriction
from above of the integration over $\bar{K}$.
The integration then is going on in the limits $[0,\bar{K}_{\rm max}]$, where
$\bar{K}_{\rm max} = \sqrt{ \bar{P}^2+ M^2 - 4m_\pi^2 }$.
On the other hand, if the measured dilepton invariant mass 
$M \le 2m_\pi$, it is necessary to introduce
a restriction on the limits of integration over the total momentum of the
lepton pair $\bar{P}$.
With keeping all this together, we write the rate of the reaction
$\pi^+\, \pi^- \rightarrow l\, \overline{l}$
in the case of a finite spatial volume of the reaction region
\begin{eqnarray}
\left \langle
\frac{ \displaystyle dR^{\rm (\rho)} }{ \displaystyle dM^2 }
\right \rangle_{\tau \to \infty}
=&&
\theta \left( M-2m_\pi \right)
\int^\infty _{-\infty} d\bar{P} \, J\left(M,R,\bar{P} \right)
\nonumber \\
\! +&& \
\theta \left( 2m_\pi-M \right)
\left[ \,
\int_{ \bar{P}_{\rm min} }^\infty d\bar{P} \, J\left(M,R,\bar{P} \right)
+
\int_{-\infty}^{ -\bar{P}_{\rm min} } d\bar{P} \, J\left(M,R,\bar{P} \right)
\right]
\, ,
\label{2-24}
\end{eqnarray}
where
$\bar{P}_{\rm min} = \sqrt{4m_\pi^2 - M^2 }$ and
\begin{eqnarray}
J\left(M,R,\bar{P} \right)
=&&
\frac{\displaystyle \alpha^2 C^2 }{\displaystyle 3(2\pi )^3 } \
\frac{ \bar{P}\, e^{-\beta \Omega(\bar{P})} }{\Omega\left( \bar{P} \right) }
\int _0^{\bar{K} _{\rm max}} \bar{K} d\bar{K}
~\frac{R}{\pi ^{1/2}} e^{ -(\bar{K} -\bar{P} )^2 R^2 } \,
\frac{\displaystyle K^2 }{\displaystyle M^2}\,
\left|F_{\pi }(K^2)\right|^2
\left( 1- \frac{4m_\pi^2}{K^2} \right) ^{3/2}
\nonumber \\
&&
\! \! \times \,
\left[
     1+\frac{1}{3M^2 K^2}
     \left(
     \Omega^2\left( \bar{P} \right)\, \left( \bar{P}-\bar{K} \right)^2
     +
     \frac{ P\cdot K }{ R^2 }
     +
     \frac{1}{ 2 R^4 }
     \right)
\right]
\label{2-25}
\end{eqnarray}
with
\begin{equation}
\Omega( \bar{P} ) = \sqrt{M^2+\bar{P}^2 } \, , \ \ \
K^2 = \Omega^2\left( \bar{P} \right)-\bar{K}^2 \, , \ \ \
P\cdot K = \Omega^2\left( \bar{P} \right) - \bar{P} \bar{K} \, , \ \
{\rm and} \ \
\bar{K}_{\rm max} = \sqrt{ \bar{P}^2+ M^2 - 4m_\pi^2 }
\, .
\label{2-26}
\end{equation}

\medskip

For completeness, let us consider the rate when we take
both limits.
Equation (\ref{2-20}) represents the rate in the case of infinite spatial
volume.
If, in addition, we consider a big enough life time of the multipion 
system, i.e. we take the limit $\tau\to \infty$,
this results in: a) appearance of the $\delta$-function on the r.h.s. of
(\ref{2-20}), i.e.
$ \lim_{\tau \to \infty} \tau \pi^{-1/2}
\exp{ \left[-\left(K_0- \Omega(\bar{K})\right)^2\tau^2\right] }
  = \delta\left(K_0 - \Omega(\bar{K})\right) $;
b) reduction of the relativistic correction factor to unity (last square
brackets on the r.h.s. of (\ref{2-20}));
c) transformation $K^2 \to M^2$.
With all these simplifications, the integration on the r.h.s. of
(\ref{2-20}) gives
\begin{eqnarray}
\left \langle
\frac{ \displaystyle dR^{\rm (\rho)} }{ \displaystyle dM^2 }
        \right \rangle_{R\to \infty, \, \tau\to \infty}
=
\frac{\displaystyle \alpha^2 C^2 }{\displaystyle 3(2\pi )^3 }
\left( 1- \frac{4m_e^2}{M^2} \right)^{1/2} \! \!
\left( 1+ \frac{2m_e^2}{M^2} \right)
\big| F_\pi(M^2) \big|^2 \,
\left( 1- \frac{4m_\pi^2}{M^2} \right)^{3/2}\,
\frac{M}{\beta} \,
K_1(\beta M)
\ .
\label{2-22}
\end{eqnarray}
So, we come to the standard result (see formula (\ref{4})) for the rate
of dilepton emission in $\pi^+ \, \pi^-$ annihilation in case of the
Boltzmann distribution of pions in the stationary and infinite
multipion system.


\subsection{ Evaluations of the dilepton emission rate }


Before going to numerically evaluate the rate of dilepton emission,
let us briefly consider the geometry of the multipion system.

After the nuclei collision, the highly excited nuclear matter goes through
several stages.
At the same time, all steps of the particle-particle transformations
take place against the background of the spatial dynamics of the system.
It is believed that the stage of hadronization and a further evolution of
the hadron system belongs to the phase of space expansion which is 
approximately isotropic for central collisions.
The latter means that the total system of pions which were created after
hadronization expands in all directions symmetrically and it is a hard
task to find even quasi-equilibrated pions in the fireball taken as a whole.
Hence, a system of pions, which are in local equilibrium, should be a small
part of the total system (fireball) where the particles move with
approximately equal radial collective velocity ${\bf v}$.
We draw this schematically in Fig.~2 where the arrows mean the radially
directed velocities of the expanding hadron matter.
Just to be transparent in consideration of finite space-time effects,
we adopt a quasi-static picture.
If, for heavy colliding nuclei, the fireball is of the mean radius
$R_0 \propto 6-10$~fm, what is known from interferometry, then the spatial
region occupied by pions which are in local equilibrium is of the size
$R \propto 1-3$~fm.
This small region is depicted in Fig.~2 as a circle of radius $R$ on
the body of the fireball.
Hence, consideration of the previous paragraphs can be attributed
to the multipion subsystem which is in the local equilibrium in a
small spatial region (on the scale of the fireball size) characterized
by radius $R$.
Evolution of this multipion subsystem in time is restricted by the mean
life time $\tau$ which is evidently also a part of the total life time
of the fireball.
\begin{figure}[t]
\begin{center}
\epsfig{file=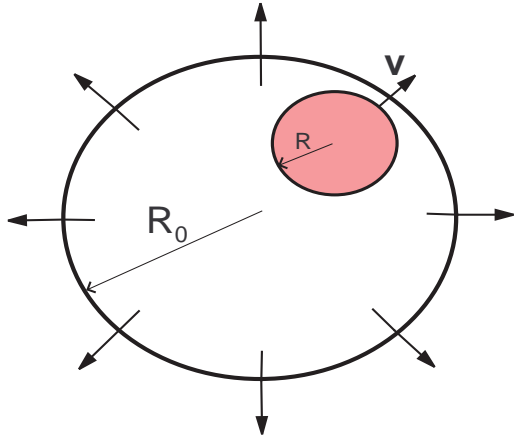,height=6cm,width=7cm,angle=0}
\caption{
The sketch of an expanding fireball.
Small circle of the radius $R$ represents the subsystem of pions
(pion gas) which is in the local equilibrium state and moves with
collective velocity ${\bf v}$.
If the fireball size is of order $R_0 \propto 6-10$~fm, then the size
of the small subsystem is  $R \propto 1-3$~fm.
  }
\vspace{-0.2cm}
\end{center}
\end{figure}

We evaluate numerically integrals (\ref{2-18}) for different
sets of parameters.
Fig.~3 shows the creation rate
$dR^{\rm (\rho)}_{e^+e^-}/dM$
of electron-positron pairs from a hot multipion subsystem
characterized by the temperature $T=180$~MeV.
Note that here and further we calculate and depicted the rate with respect
to the invariant mass $M$, but not $M^2$ (to obtain this, we multiply the r.h.s.
of Eq.~(\ref{2-18}) by $2M$).
There are seven curves in this figure which are labeled by numbers
$1, \ldots ,7$ in the following correspondence to the values of the
mean size of the pion subsystem and its life time:
$1)$ $R=1$~fm, $\tau=1$~fm/c;
$2)$ $R=2$~fm, $\tau=2$~fm/c;
$3)$ $R=3$~fm, $\tau=3$~fm/c;
$4)$ $R=5$~fm, $\tau=5$~fm/c;
$5)$ $R=10$~fm, $\tau=10$~fm/c;
$6)$ $R=20$~fm, $\tau=20$~fm/c;
$7)$ $R=\infty$, $\tau=\infty$.
Evaluation for the infinite space-time volume, which corresponds to
curve $7$ in Fig.~3, repeats the standard result, as was shown
in (\ref{2-22}).
Notice how the $e^+e^-$ emission rate deviates from the standard one (curve 7):
it is finite in the region of the lepton invariant mass
$M\le 2m_\pi$.
Qualitatively, as we discussed above, it is a consequence of the
uncertainty principle.
It is of interest to note that the deviation of the rate for a restricted 
reaction zone from the standard one calculated for the infinite space-time 
volume (curve 7) is more sizable in the region of small invariant masses.
This seems natural because the uncertainty (quantum fluctuations) is more
pronounced for smaller energies (masses).
We emphasize as well that the tendency of curves 1, 2
in Fig.~3 is similar to CERES data.

In Fig.~4, we show the evaluation of the creation rate
$dR^{\rm (\rho)}_{\mu^+\mu^-}/dM$
for $\mu^+\mu^-$-pairs at the same temperature $T=180$~MeV, as for
electron-positron creation.
There are also seven curves in this figure which are labeled by numbers
from $1$ to $7$ in the same correspondence to the values of the
mean size $R$ and mean life time $\tau$ of the pion subsystem as we take
for the previous figure.
Notice that the $\mu^+\mu^-$
creation rate for a finite pion system deviates from the rate for an
infinite pion system as bigger as smaller the spatial size and life time
of the pion system in the same way as for $e^+e^-$ creation.
Of course, it is a reflection of the uncertainty principle and broken 
translation
invariance which is expressed by the presence of the distribution
$|\rho(K-P)|^2$ as the integrand factor in (\ref{2-15a}).
Technically, the presence of the Gaussian form-factor of the pion
subsystem (\ref{2-15}) results in a broadening of the rate which is
as wider as smaller the parameters $R$ and $\tau$ of the Gaussian exponents
$\exp{[-(\bar{K} -\bar{P} )^2 R^2]}$ and
$\exp{ [-(K_0 - P_0)^2\tau^2] }$
in the integrand on the r.h.s. of (\ref{2-18}).
Evidently, the biggest broadening was obtained for the pion subsystem
with mean radius $R=1$~fm and life time $\tau=1$~fm/c.
\noindent
\begin{figure}[t]
\begin{center}
\epsfig{file=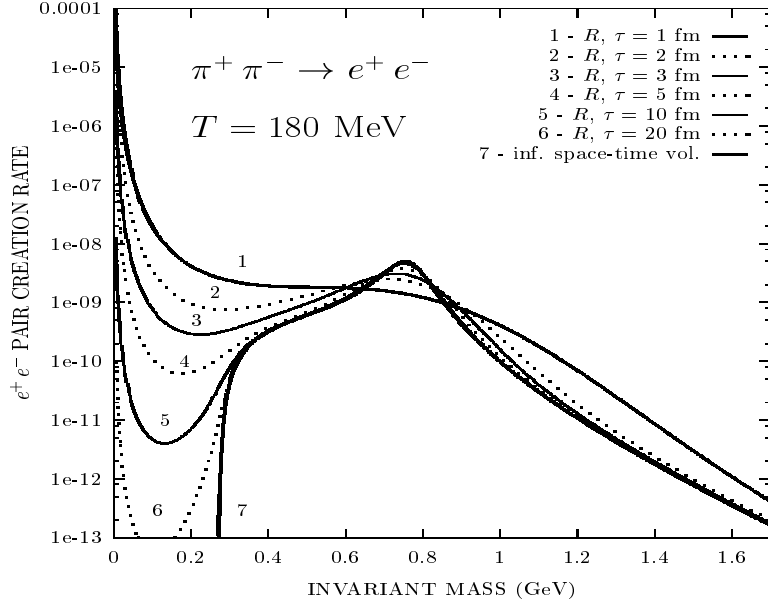,height=8cm,width=10cm,angle=0}
\noindent
\caption{
The rate $dR^{\rm (\rho)}_{e^+e^-}/dM$ of electron-positron pair
creation in pion-pion annihilation.
Different curves correspond to the different spatial sizes $R$
and different life times $\tau$ of a hot pion system, $T=180$~MeV.
  }
\vspace{-0.2cm}
\end{center}
\end{figure}
\noindent
\begin{figure}[t]
\begin{center}
\epsfig{file=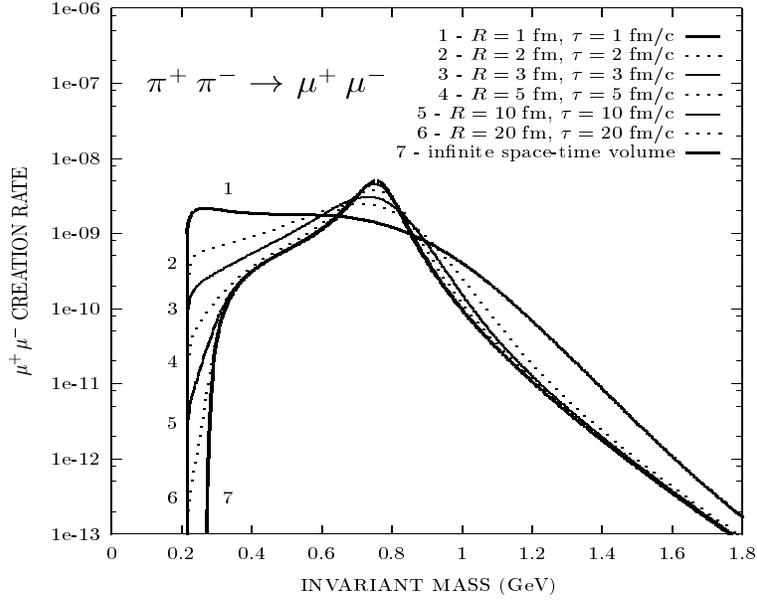,height=8cm,width=10cm,angle=0}
\noindent
\caption{
The rate $dR^{\rm (\rho)}_{\mu^+\mu^-}/dM$ of $\mu^+\mu^-$ pair creation
in pion-pion annihilation.
Different curves correspond to the different spatial sizes $R$
and different life times $\tau$ of a hot pion system,
$T=180$~MeV.
  }
\vspace{-0.2cm}
\end{center}
\end{figure}
\noindent
Fig.~4 shows in contrast to the standard result that the
threshold of $\mu^+\mu^-$ creation rate is at the value of invariant mass
$M_{\rm thresh}^{\mu^+\mu^-}=2m_\mu \approx 211.3$~MeV.
An analogous threshold for electron-positron creation, which is
$M_{\rm thresh}^{e^+e^-}=2m_e \approx 1.02$~MeV,
does not visible on the scale of mass span which is taken in Fig.~3.
In Fig.~5, we show the rate $dR^{\rm (\rho)}_{e^+e^-}/dM$ of
electron-positron pair creation in the region of threshold for the
mean radius $R=1$~fm and life time $\tau=1$~fm/c of the pion
subsystem.
Solid curve is a result of evaluation of expression (\ref{2-18}).
The dashed curve 2 is a result of evaluation of expression (\ref{2-18}) 
which is shorted by dropping out the last square brackets on the r.h.s. 
of the equation.
\noindent
\begin{figure}[t]
\begin{center}
\rotatebox{-90}{
\epsfig{file=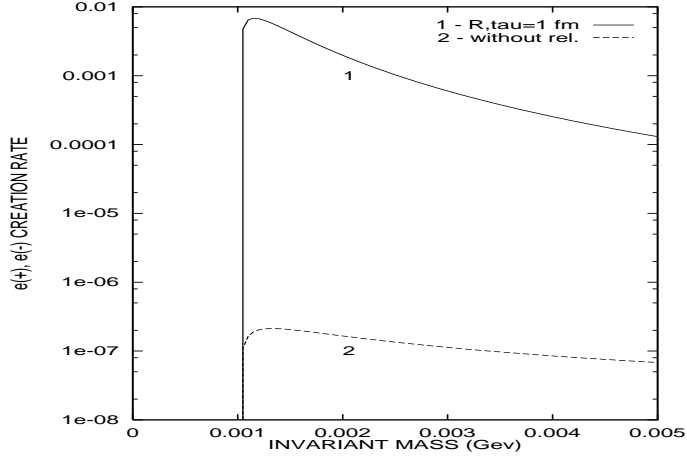,height=9cm,width=6cm,angle=0}
}
\noindent
\caption{
The rate $dR^{\rm (\rho)}_{e^+e^-}/dM$ of electron-positron pair
creation in pion-pion annihilation in the threshold region.
Mean spatial size $R=1$~fm,
life time $\tau=1$~fm/c,
$T=180$~MeV.
Curve 1 is evaluated in accordance with expression (\ref{2-18}).
Curve 2 is evaluated for the same integral (\ref{2-18}) but without
the last factor in square brackets which reflects a relative movement of
the pion-pion c.m.s. and the electron-positron c.m.s.
  }
\vspace{-0.2cm}
\end{center}
\end{figure}
\noindent
We recall that the expression in these square brackets carries the effect of
a relative movement of the pion-pion c.m.s. and the electron-positron c.m.s. which
is a consequence of the finiteness of the pion-pion reaction region.
We see that, in the region of the electron-positron creation
threshold $M=2m_e$, this effect (curve 1) is realized as four additional
orders to the creation rate if one does not take
into account the relative velocity of the
$\pi^+\, \pi^-$ c.m.s. and the $e^+\, e^-$ c.m.s. (curve 2).

Fig.~6 shows the same evaluations as the previous one but in a wider range
of invariant masses.
Curves 1, 2, 3 are evaluated in accordance
with formula (\ref{2-18}), whereas, curves 1n, 2n, 3n are evaluated in
accordance with Eq.~(\ref{2-18}) where the relativistic
factor (the last square brackets on the r.h.s. of (\ref{2-18}))
is dropped out.
The dash-dotted curves 1 and 1n correspond to
$R=1$~fm, $\tau=1$~fm/c;
the solid curves 2 and 2n correspond to
$R=3$~fm, $\tau=3$~fm/c;
the dashed curves 3 and 3n correspond to
$R=10$~fm, $\tau=10$~fm/c,
and the long-dash-dotted curve 4 corresponds to the
infinite space-time volume of the reaction region.
We see that the relativistic effect under consideration is attributed
to the region of invariant masses which is below the invariant mass
$M=2m_\pi$ and it increases with decreas of the size of the pion system.
The latter is partially  explained by the presence of
last two terms $P\cdot K/R^2$ and $1/(2R^4)$ in square brackets on the
r.h.s. of (\ref{2-18}).
We do not present the same figure for muon pair creation because this
effect is practically not visible even for the pion system of the size
$R=1$~fm, $\tau=1$~fm/c because of the close position of the real threshold
which equal $2m_\mu$ to the standard one which equals $2m_\pi$.
\begin{figure}[t]
\begin{center}
\rotatebox{-90}{
\epsfig{file=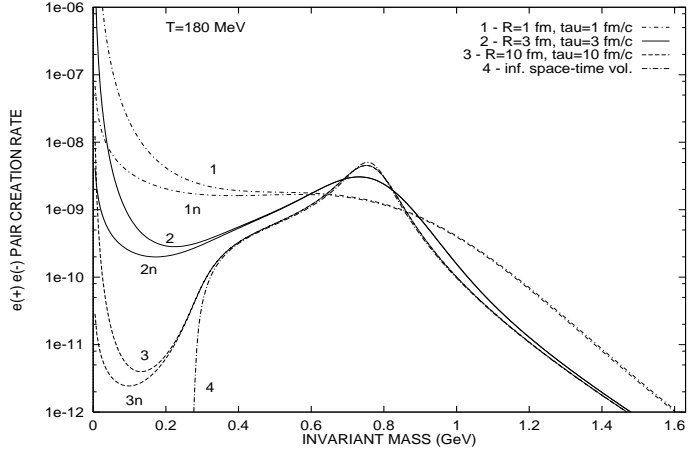,height=9cm,width=6cm,angle=0}
}
\caption{
The rate $dR^{\rm (\rho)}_{e^+e^-}/dM$ of electron-positron pair
creation in pion-pion
annihilation with (curves 1, 2, 3) and without
(curves 1n, 2n, 3n) a c.m.s. relativistic correction.
Different curves correspond to the different spatial sizes $R$
and different life times $\tau$ of the pion system.
Curve 4 corresponds to the infinite size of the pion system.
  }
\vspace{-0.2cm}
\end{center}
\end{figure}
\noindent

Next, we compare the emission rates obtained in the frame of two models (a)
and (b).
Model (a) (Fig.~1a) reflects the presence of a dense hadron environment
which affects the $\rho$-meson properties.
Due to the medium effects, the $\rho$-meson mean free path and mean life time
are strongly suppressed inside the dense nuclear matter.
As a limit of this suppression,  we adopt a zero mean free path and zero 
life time in model (a) .
At the same time, to preserve a $\rho $-meson nature of $\pi^+\, \pi^-$
annihilation, we keep the $\rho $-meson form-factor
$|F_\pi (M^2)|^2$ (\ref{3}) with vacuum parameters in model (a).

Model (b) (Fig.~1b) is a standard representation of the vector
meson dominance.
The idea was to formulate this model as a contra-pole to model (a) -
the $\rho $-meson contribution to the process
$\pi^+\, \pi^- \to  \rho \to \gamma^* \to l\, \overline{l}$
is not affected at all by the hadron environment.
Then the real picture of the pion annihilation in the dense nuclear matter
will be somewhere in between these two limit models.

In fact, a more accurate picture, if we assume the transformation of
a $\rho$-meson to a photon $\gamma^*$ inside the fireball, should deal
with $\rho$-meson propagator as in model (b) but with the vertex $x_1'$ which
belongs to the restricted volume of the fireball.
Then, the integration over $x_1'$ will be weighted by one additional
distribution $\rho(x_1')$ as the integrand factor in amplitude (\ref{2-1}).
In a sense, this complication will not bring a qualitatively new result.
\begin{figure}[t]
\begin{center}
\rotatebox{-90}{
\epsfig{file=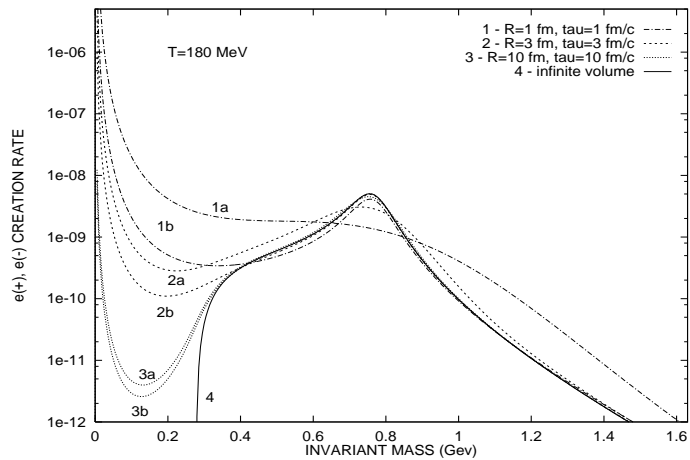,height=9cm,width=6cm,angle=0}
}
\noindent
\caption{
The rate $dR^{\rm (\rho)}_{e^+e^-}/dM$ of electron-positron pair
creation in pion-pion annihilation.
All curves which are labeled by letter "a" were evaluated in the
model which  corresponds to the diagram depicted in Fig.~1a.
All curves which are labeled by letter "b" were evaluated in the
model which  corresponds to the diagram depicted in Fig.~1b.
The values of the mean spatial size and life time:
dash-dotted curves (1a and 1b) - $R=1$~fm, $\tau=1$~fm/c;
dashed curves (2a and 2b) - $R=3$~fm, $\tau=3$~fm/c;
dotted curves (3a and 3b) - $R=10$~fm, $\tau=10$~fm/c;
solid curve (4) - infinite space-time volume.
Effective temperature $T=180$~MeV.
  }
\vspace{-0.2cm}
\end{center}
\end{figure}

Fig.~7 shows the rate $dR^{\rm (\rho)}_{e^+e^-}/dM$ of electron-positron
pair creation evaluated in the frame of models (a) and (b).
Curves 1a, 2a, 3a are obtained as before by the evaluation of
formula (\ref{2-18}).
Curves 1b, 2b, 3b are obtained by the evaluation of formula (\ref{2-18}) 
when the argument of the pion form-factor is changed from $K^2$ to $M^2$.
Then it can be written in front of the integral
in the form
$\big| F_\pi(M^2) \big|^2$.
This change corresponds to the integration of the vertex $x_1'$
(see Fig.~1a) over infinite space-time volume.
The dash-dotted curves 1a and 1b correspond to $R=1$~fm, $\tau=1$~fm/c;
the dashed curves 2a and 2b correspond to $R=3$~fm, $\tau=3$~fm/c;
the dotted curves 3a and 3b correspond to $R=10$~fm, $\tau=10$~fm/c,
and the solid curve 4 corresponds to the infinite space-time volume of the 
reaction region.
The temperature of the locally equilibrated pion system was taken as
$T=180$~MeV.
Comparison of curves 1a and 1b evidently shows that the $\rho$-meson
peak becomes much more pronounced when we shift from model (a) to
model (b).
Hence, we can conclude, that due to the small mean size and life time of the
pion subsystem, what results in the broad enough pion system form-factor
$\rho(K-P)$, in the frame of model (a), we obtain a strong smearing of the
$\rho$-meson form-factor $\left| F_\pi(K^2) \right| ^2$ (see Eq.~(\ref{3})).
But this difference becomes smaller with increasing the mean parameters 
$R$ and $\tau$ of the pion system. 
For instance, the behavior of curves 2a and 2b is qualitatively the same
for $R=3$~fm, $\tau=3$~fm/c.

Fig.~8 shows the emission rate $dR^{\rm (\rho)}_{\mu^+\mu^-}/dM$ of the
$\mu^+\, \mu^-$ pair creation for models (a) and (b).
We see that model (a) (dash-dotted curve 1a) again reveals indeed a very
strong smearing of the $\rho$-meson peak because of the presence of the
pion system form-factor $\rho(K-P)$ when it characterized by the smallest
parameters under consideration: $R=1$~fm, $\tau=1$~fm/c.
Meanwhile, for the same set of parameters, the $\mu^+\, \mu^-$ emission 
rate evaluated in the frame of model (b)
\noindent
\begin{figure}[t]
\begin{center}
\rotatebox{-90}{
\epsfig{file=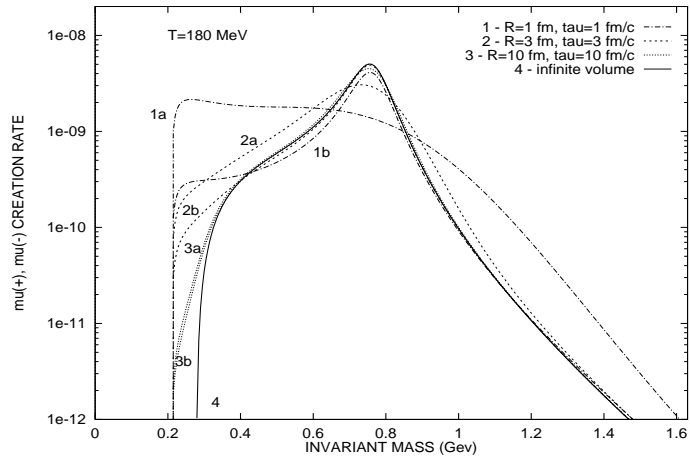,height=9cm,width=6cm,angle=0}
}
\noindent
\caption{
The rate $dR^{\rm (\rho)}_{\mu^+\mu^-}/dM$ of $\mu^+\, \mu^-$ pair
creation in pion-pion annihilation.
All curves are labeled as in Fig.~7.
The set of parameters is also taken as in Fig.~7.
  }
\vspace{-0.2cm}
\end{center}
\end{figure}
\noindent
(dash-dotted curve 1b)
practically coincides with the standard result (solid curve 4) in the
region $M \ge 450$~MeV.
But, for invariant masses $M \le 450$~MeV, the emission rate reveals a
strongly pronounced flat shoulder with a cut off at the threshold
$M=2m_\mu=211.3$~MeV.
For the parameters $R=3$~fm, $\tau=3$~fm/c, the emission rates for
models (a) and (b) are not so distinguishable for invariant masses above 
the $\rho$-peak (dotted curves 2a and 2b).
But when the invariant mass is below the $\rho$-peak, these emission rates 
manifest a well visible corridor where the real rate, as we discussed before, 
should be found.
At last, when the mean size of the pion subsystem is $R=10$~fm, i.e. is 
comparable to the size of the fireball, and the mean life time is
$\tau=10$~fm/c, which is also comparable with the life time of the fireball,
the emission rates evaluated in the frame of models (a) and (b) are not
distinguishable (dense dotted curves 3a and 3b), but a cut off of the rates
is at the threshold $M=2m_\mu$.

\medskip


\section{Discussion and conclusion}


In the present work, we have studied the effect of pion-pion annihilation
in a finite space-time volume on the dilepton yield.
Due to the uncertainty principle, a restriction of the volume and time of
the reaction
$\pi^+\, \pi^- \to l^+\, l^-$ results
in breaking the detailed energy-momentum conservation in the s-channel.
Formally, this expresses in the alteration of
$(2\pi)^4\delta ^4(K-P)$ to the distribution $\rho(K-P)$ (or the 
form-factor of the multipion subsystem),
where $K=k_1+k_2$ is the total 4-momentum of the annihilating pion pair
and $P=p_++p_-$ is the total 4-momentum of the created lepton pair.
For this kinematics, the energy-momentum conservation is valid as an integral
law which is guaranteed by relation
$\int d^4K \rho(K-P)=(2\pi)^4$, where the registered 4-momentum
$P$ plays a role of the center (or mean value) of the distribution $\rho(K-P)$
with respect to the total 4-momentum $K$ of the annihilating pair as a
fluctuate quantity.
As a result of the above mechanism, the rate of the lepton pair yield
is a finite quantity below the threshold $M=2m_\pi$.
Actually, the real threshold is determined as the total mass of the
registered particles, i.e. for  $e^+\, e^-$ production (see Figs.~4 and 6)
$M_{\rm thresh}^{e^+e^-}=2m_e$,
where $m_e$ is the electron mass, and for  $\mu^+\, \mu^-$ production (see Fig.~4)
$M_{\rm thresh}^{\mu^+\mu^-}=2m_\mu$, where $m_\mu$ is the mass of
$\mu$-meson.
Note that, in the calculations made, we kept the invariant mass of the 
pion-pion pair not less than $2m_\pi$, or the range of integration over the
total 4-momentum of the pion pair $K$ was determined as $2m_\pi \le K^2$.

To explain technically why is that, we recall that replacing a $\delta$-function
in the integrand by a distribution $\rho(K - P)$ (for instance in
(\ref{5})) results in the additional integration where the invariant mass $M$
(or $P^0$ and ${\bf P}$) plays a role of external parameter.
Then, if the distribution $\rho(K - P)$ is wide enough, it can lead to
the finite results of integration for those values of $M$ which were not
in game before.
To give intuitive feeling how it is going on, let us consider the toy averaging
of a "good" function
$ R(M') $
over the variable $M'$ by using the distribution function
$ \rho(M'-M)= \tau\, \exp{[-(M'-M)^2 \tau^2] } /\sqrt{\pi} $
which has the mean value $M$ (measurable quantity) and the width of the
distribution $1/\tau$.
Then, this averaging reads
\begin{equation}
\langle R\, \rangle (M) =
\int _{2m_\pi}^\infty dM'~R(M')\,
\frac{\tau}{ \sqrt{\pi} } \,
e^{ -\left( M' - M\right)^2 \tau^2 }
\ ,
\label{4a}
\end{equation}
where the variable $M'$ (analog of the two-pion invariant mass) varies in
the region
$ 2m_\pi < M' < \infty $.
For long enough life-times ($\tau \to \infty$), we immediately obtain, as
a result of averaging the function $R$ taken at the value $M'=M$, i.e.
$\langle R \rangle (M) = R(M) \theta(M-2m_\pi) $.
The presence of the $\theta$-function in this expression stresses
the fact that, for the infinite life time, the invariant mass $M$ of a lepton
pair should be bigger than the
total mass of annihilating particles, it is $2m_\pi$ in the case
under consideration.
Meanwhile, for a finite life time, for instance $\tau =1--4\,$fm/c, the
integral is not zero for mean values of the distribution $M$ which are
less than $2m_\pi$. 
For example, we can consider even the extremely small magnitude 
$M=2 m_e \approx 1$~MeV.
Then, though the distribution function $\rho(M'-M)$ is centered
around the mean value $M=2 m_e$, i.e. beyond the region of integration
$ 2m_\pi \leq M' < \infty $, for small life times $\tau$, the right wing of
the distribution $\rho(M'-M)$ is valued enough to make the result of
integration on the r.h.s. of (\ref{4a}) quite significant to be taken into
account.
The illustration of this assertion is depicted in Fig.~9, where the integrand
in (\ref{4a}) is shown schematically as solid line.
The value of integral (\ref{4a}) $\langle R\, \rangle (M)$ equals
the shaded area which starts from the value $M'=2m_\pi$.
\begin{figure}[t]
\begin{center}
\epsfig{file=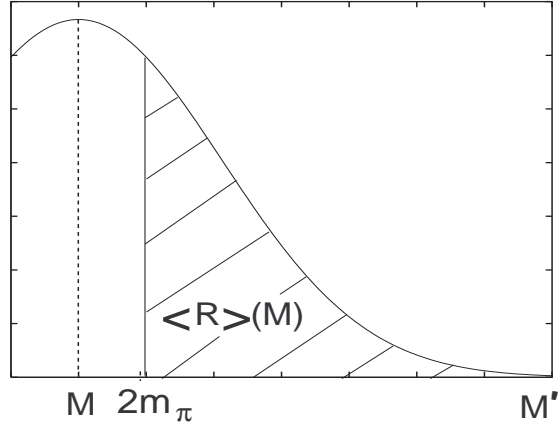,height=6cm,width=8cm,angle=0}
\caption{ Sketch of the integrand in Eq.~(\ref{4a}).
The integral is evaluated when the value of the parameter $M$ is less
than the lower limit of integration, $2m_\pi$.
  }
\vspace{-0.5cm}
\end{center}
\end{figure}
\noindent
In a similar way, if the life time of the pion gas is small enough, 
we obtain a nonvanishing rate for dilepton invariant masses which 
are less than $2m_\pi$,
whereas the invariant mass of the two-pion system (it is $M'$ in
our toy example) is certainly bigger than $2m_\pi$.

One of the consequences of breaking the exact equality between
the total 4-momentum $K$ of an incoming pion pair and
the total 4-momentum $P$ of an outgoing lepton pair is a noncoincidence
of two center-of-mass systems:
the $\pi^+\, \pi^-$ c.m.s. differs from the $l^+\, l^-$ c.m.s.
For the technics of calculations, this means the following:
if some quantity was obtained as the result of calculations
in the $l^+\, l^-$ c.m.s. and we need it in the $\pi^+\, \pi^-$ c.m.s.,
we should make a Lorentz transformation from the system which moves with
velocity ${\bf v}_P={\bf P}/P_0$ in the lab system to a system which
moves in the lab system with velocity
${\bf v}_K={\bf K}/K_0$ for every particular value $P$ and $K$;
then this transformation will be integrated with respect to $K$ exploiting
the distribution $\rho(K-P)$.
The last procedure brings an additional deviation of the dilepton yield
from the standard one (see Figs.~5 and 7).
As seen from Fig.~6, the manifestation of this effect for
$e^+\, e^-$ production is essential in the region which is below $M=2m_\pi$
and the effect is washed out practically for invariant masses which are 
bigger than two pion masses.
As we showed in Fig.~4 near the real threshold
$M_{\rm thresh}^{e^+e^-}=2m_e$, this effect gives an additional sizable
contribution to the $e^+\, e^-$ yield increasing the rate up to four orders.
Note that, this effect is not visible practically for $\mu^+\, \mu^-$ 
production (see Fig.~4) because the threshold invariant mass is big enough,
$M_{\rm thresh}^{\mu^+\mu^-}=2m_\mu \approx 211.3$~MeV, and lies very close
to $M=2m_\pi$.
Basically, the manifestation of quantum fluctuations due to a finite size of 
the reaction zone is not so strong for $\mu^+\, \mu^-$ production as for
$e^+\, e^-$ one just because of the big threshold mass.

We use the concept that a dense hadron medium suppresses the
$\rho$-meson mean free path and life time.
As a limit of this suppression, one can assume a zero mean free path and
zero mean life time of a $\rho$-meson.
We considered two models of pion-pion annihilation when pions are confined
to the finite hadron medium.
The first model (a) (Fig.~1a) mimics an extreme influence of dense hadron
environment through adopting a $\rho$-meson mean free path and mean life
time to be equal to zero,
whereas the $\rho$-meson form-factor is taken as the vacuum one.
The second model (b) (Fig.~1b) deals with the standard concept of vector
meson dominance and exploits the vacuum  $\rho$-meson form-factor.
The latter means the absence of any suppression due to the environment.
By this, the models under consideration reflect two limit cases -
an extreme influence of dense hadron medium (model (a)) and the absence of
influence of hadron environment (model (b)) on the $\rho$-meson mean
free path and mean life time.
Then, the evaluations of dilepton rates in the frame of both models mean
a determination of the corridor which should absorb a real rate for any
kind of hadron medium.
We compare the rates evaluated in the frame of both models.
Fig.~7 shows the rates $dR^{\rm (\rho)}_{e^+e^-}/dM$ of the
electron-positron pair emission for models (a) and (b), and
Fig.~8 shows the rates $dR^{\rm (\rho)}_{\mu^+\mu^-}/dM$ of the
$\mu^+\mu^-$ pair emission for models (a) and (b) as well.
We found that the difference of curves (a) and (b) is sizable just for
small sizes of the multipion systems, e.g.
from $R=1$~fm, $\tau=1$~fm/c to $R=3$~fm, $\tau=3$~fm/c.
But, with increase of the size, it is seen the strong convergence of the rates
(a) and (b).
Indeed, when the parameters of the multipion system are around the
values $R=10$~fm, $\tau=10$~fm/c, there is practically no difference
between curves (3a) and (3b) in Figs.~7,8 for $e^+\, e^-$ 
and $\mu^+\, \mu^-$ productions.

So, if annihilating (interacting) particles are confined to a finite
space-time volume, this inspires several mechanisms which give rise
to the new features of the dilepton production rate of the process
$\pi^+\, \pi^- \to l^+\, l^-$ in comparison to the standard rate which is
attributed to the infinite space-time.
In the present paper, we considered just two of these mechanisms:
1) Direct contribution from the multiparticle (multipion) system form-factor
and
2) The contribution from the relativistic transformation of quantities from
the c.m.s. of outgoing particles ($l^+\, l^-$ pair) to the c.m.s. of incoming
particles ($\pi^+\, \pi^-$ pair).
It is found that even these two contributions give sizable effects
on the dilepton yield.
Meanwhile, the third contribution which comes from a nonstationarity of
the confined multiparticle (multipion) system is still left beyond the
scope of our paper.
We reserve a consideration of this problem to the next paper which is in
progress as well as an application of the approach elaborated to other
reactions which take place in quark-gluon plasma and hadron plasma
confined to a finite volume, for instance, $q\bar{q} \to l\bar{l}$, etc.

\section*{Acknowledgements}

One of the authors (D.A.) would like to express his gratitude for warm
hospitality to the staff
of the Physical Department of the University of Jyv\"askyl\"a where this
work was started and mainly done. 
D.A. is thankful to  U.~Heinz,  J.~Pisut and  
 E.~Suhonen for discussions and support.


\appendix


\section{ } 
\subsection*{ Kinetic description of the process 
$\pi^+\, \pi^- \to  l\, \overline{l}$
  }


Let us summarize the standard results which were obtained for the process
$\pi^+\, \pi^- \to  \rho \to \gamma^* \to l\, \overline{l}$
within the kinetic approach \cite{gale87} (see also
\cite{mclerran85,domokos,ruuskanen,kapusta89}).
The two-pion annihilation rate in the infinite space-time volume,
is obtained within the kinetic approach
by integration over pion momenta ${\bf k}_1$ and ${\bf k}_2$
which are weighted by the pion distribution function $f(E)$
(with $ E=\sqrt{ m_\pi^2 + {\bf k}^2 } $),
\begin{equation}
\frac{\displaystyle dN}{\displaystyle dx^4dM^2}=
\int \frac{d^3 k_1}{(2\pi )^3}f(E_1)
\int \frac{d^3 k_2}{(2\pi )^3}f(E_2)\, v_{12}\, \sigma _{\pi \pi }(M) \,
\delta \left[ (k_1+k_2)^2-M^2\right] \ ,
\label{1}
\end{equation}
where $dx^4$ is an element of the 4-volume,
$v_{12}=\sqrt{(k_1\cdot k_2)^2 -m_\pi^4 }/E_1E_2 $
is the pion-pion relative velocity, $M$ is the invariant mass of the 
dilepton system.
The pion annihilation cross section is obtained by integration over lepton
momenta
${\bf p}_+$ and ${\bf p}_-$ keeping the equality of the total pion-pion
energy-momentum $K=k_1+k_2$ and total dilepton energy-momentum $P=p_++p_-$
by the $\delta$-function:
\begin{equation}
\sigma _{\pi \pi }(M)=\frac{1}{J}
\int \frac{d^3p_+}{(2\pi )^3\, 2E_+}\, \frac{d^3p_-}{(2\pi )^3\, 2E_-} \,
\frac{|m_{fi}|^2\, \left| F_{\pi }(M^2) \right|^2 }
     {2E_1\, 2E_2\, (k_1+k_2)^4 }
\, \delta ^4(k_1 + k_2 - p_+ - p_-)
\ ,
\label{5}
\end{equation}
where
$J=v_{12}=\sqrt{(k_1\cdot k_2)^2 -m_{\pi }^4}/E_1\, E_2$
is the pion-pion relative current, and $m_{fi}$ is the matrix element of 
the process, 
$ | m_{fi} | ^2 = \left[ 8(k_{1}-k_2)\cdot p_+\
(k_1 - k_2)\cdot p_- -4(k_1-k_2)^2 p_+\cdot p_- \right] $.
Evaluation of the integral on the r.h.s. of (\ref{5}) results in
\begin{equation}
\sigma_{\pi \pi }(M)=\frac{4\pi }{3} \frac{\alpha^2}{M^2}
\sqrt{1-\frac{4m_\pi^2}{M^2} } \
\left| F_\pi (M^2) \right| ^2 \ ,
\label{2}
\end{equation}
where the $\delta$-function on the r.h.s. of (\ref{1})
was taken into account by putting $(k_1+k_2)^2=M^2$.
The $\rho $-meson form-factor $F_\pi (M)$ is taken as
(see \cite{gale87,gounaris})
\begin{equation}
\left| F_\pi(M^2) \right| ^2 =
\frac{ m_\rho^4 }{ (M^2-{m'_\rho}^2)^2
  +m_\rho^2 \, \Gamma_\rho^2 } \ ,
\label{3}
\end{equation}
and
$m_\rho=775\, {\rm MeV}, \  m'_\rho=761\, {\rm MeV}, \
\Gamma_\rho=118\, {\rm MeV}$.

If the pion momentum-space distribution function is taken as the Boltzmann 
one,
$ f(E)=\exp{(-E/T)} $, where $T$ is a temperature of the pion gas, then
it is straightforward to evaluate the annihilation rate (\ref{1}):
\begin{eqnarray}
\frac{\displaystyle dN}{\displaystyle dx^4dM^2} =
\frac{ \alpha^2 }{ 3(2\pi )^3 } M\, T\,
K_1\left( \frac{M}{T} \right)\, \left| F_\pi(M^2) \right|^2
\left( 1-\frac{ 4m_\pi^2 }{ M^2 } \right)^{3/2}
\ .
\label{4}
\end{eqnarray}
It is necessary to point out that the
$\delta$-function in (\ref{5}) which expresses a "sharp" energy-momentum
conservation in the s-channel of the reaction
$ \pi^+\, \pi^- \to  \rho \to \gamma^* \to l\, \overline{l} $
appears due to the approximation where the space-time volume of the 
multipion system is taken as infinite one.
As a consequence of the equality $k_1+k_2=p_+ +p_-$, which is brought by
the $\delta$-function in (\ref{5}), the $\delta$-function on the r.h.s. 
of (\ref{1}) indicates that the invariant mass of the
two-pion system is the same, namely $(k_1+k_2)^2=M^2$, as the invariant 
mass $M$ of dilepton system ($(p_+ + p_- )^2=M^2$), which
is a measurable quantity.


\section{}
\subsection*{ Transformation of 
             $ \left( {\bf k}_1-{\bf k}_2 \right)_P^2 $
             from the P-system to the K-system }

We start from the covariant relation
\begin{equation}
q_P^2 = q_K^2
\ ,
\label{a11}
\end{equation}
where we adopt the shorthand notation $q=k_1-k_2$.
Subindex $P$ or $K$ denotes a particular realization of the quantity
$q^2$ in the P-system (lepton pair c.m.s.) or in the K-system (pion pair c.m.s.).
Because $q^0_K=k^0_1-k^0_2=0$ in the K-system, one can rewrite (\ref{a11})
in the form
\begin{equation}
{\bf q}_P^2 = {\bf q}_K^2 + \left( q_P^0 \right)^2
\ .
\label{a12}
\end{equation}
On the other hand, we write $P^2=P_0^2$ in the P-system where 
$P=(P_0,{\bf P}=0)$ and hence obtain
$ \left( q_P^0 \right)^2 =  (q\cdot P)^2/P^2 $.
As a next step, we write this expression in the K-system
\begin{equation}
 \frac{1}{P^2} (q\cdot P)^2 =
 \frac{1}{P^2} (q_K\cdot P_K)^2 =
 \frac{1}{P^2}\, {\bf q}_K^2 \, {\bf P}_K^2 \cos^2{\theta}
\ ,
\label{a14}
\end{equation}
where $\theta $ is the angle between vectors ${\bf q}_K$
and ${\bf P}_K$ (both are in the K-system).
Putting everything together, we obtain
\begin{equation}
{\bf q}_P^2 = {\bf q}_K^2
\left( 1+ \frac{{\bf P}_K^2}{P^2} \cos^2{\theta} \right)
\ .
\label{a15}
\end{equation}

We need to do the last step of the transformations.
It is necessary to transform the quantity ${\bf P}^2$ from the lab system
to the K-system as we need it in (\ref{a15}).
For this transformation, let us exploit the covariance of the product
$K\cdot P$ (as we did it before with $q\cdot P$).
Indeed, we write $K^2=K_0^2$ in the K-system where $K=(K_0,{\bf K}=0)$ and 
hence obtain $ K^2\left( P_K^0 \right)^2 =  (P\cdot K)^2 $.
Taking this into account, one can immediately find
\begin{equation}
{\bf P}_K^2 = \left( P_K^0 \right)^2 - P^2
= \frac{(P\cdot K)^2}{K^2} -P^2
\ .
\label{a16}
\end{equation}
Putting this into (\ref{a15}), we come to the required result
\begin{equation}
\left( {\bf k}_1 - {\bf k}_2 \right)_P^2 =
\left( {\bf k}_1 - {\bf k}_2 \right)_K^2
\left[ 1+ \left( \frac{(P\cdot K)^2}{P^2\, K^2} - 1 \right) \cos^2{\theta}
\right]
\ ,
\label{a17}
\end{equation}
where $\theta $ is the angle between vectors ${\bf k}_1 - {\bf k}_2$
and ${\bf P}$ when they both are considered in the pion pair c.m.s. 
(K-system).
So, the expression on the r.h.s. of (\ref{a17}) is related to the 
pion pair c.m.s.

\end{document}